\documentclass[a4paper,11pt]{article} 
\usepackage{pos}
\title{LOgging UnifieD for ASTRI Mini Array}
\ShortTitle{LOUD system for ASTRI Mini Array}
\author*[a,1]{Federico Incardona}
\author[a,1]{Alessandro Costa}
\author[a,1]{Kevin Munari}
\author[a,1]{Pietro Bruno}
\author[d,1]{Andrea Bulgarelli}
\author[b,1]{Stefano Germani}
\author[a,1]{Alessandro Grillo}
\author[c,1]{Joseph Schwarz}
\author[a,1]{Eva Sciacca}
\author[b,1]{Gino Tosti}
\author[e]{Fabio Vitello}
\author[a]{Giuseppe Tudisco}
\note{for the ASTRI Project:  \url{http://www.astri.inaf.it/}}
\affiliation[a]{INAF, Osservatorio Astrofisico di Catania, Via S Sofia 78, I-95123 Catania, ITALY}
\affiliation[b]{Universit\`a di Perugia, Dipartimento di Fisica e Geologia, IT}
\affiliation[c]{INAF, Osservatorio Astronomico di Brera, IT}
\affiliation[d]{INAF, Osservatorio di Astrofisica e Scienza dello Spazio di Bologna, IT}
\affiliation[e]{INAF, Istituto di Radiastronomia, Bologna, IT}
\emailAdd{federico.incardona@inaf.it}
\abstract{The ASTRI (Astrofisica con Specchi a Tecnologia Replicante Italiana) Mini-Array (MA) project is an international collaboration led by the Italian National Institute for Astrophysics (INAF). ASTRI MA is composed of nine Cherenkov telescopes operating in the energy range 1-100 TeV, and it aims to study very high-energy gamma ray astrophysics and optical intensity interferometry of bright stars. ASTRI MA is currently under construction, and will be installed at the site of the Teide Observatory in Tenerife (Spain). The hardware and software system that is responsible of monitoring and controlling all the operations carried out at the ASTRI MA site is the Supervision Control and Data Acquisition (SCADA). The LOgging UnifieD (LOUD) subsystem is one of the main components of SCADA. It provides the service responsible for collecting, filtering, exposing and storing log events collected by all the array elements (telescopes, LIDAR, devices, etc.). In this paper, we present the LOUD architecture and the software stack explicitly designed for distributed computing environments exploiting Internet of Things technologies (IoT).}
\FullConference{37$^{\rm{th}}$ International Cosmic Ray Conference (ICRC 2021)\\
		July 12th -- 23rd, 2021\\
		Online -- Berlin, Germany}
\begin{document} 
\maketitle
%\keywords{astri, astri-ma,  monitoring, logging, alarms, Cherenkov Telescope, CTA}
%
%
\section{Introduction}
The ASTRI (Astrofisica con Specchi a Tecnologia Replicante Italiana) Mini-Array (MA) \cite{antonelliLA} will be a gamma-ray telescope array operating in the energy range 1-100 TeV and beyond from the Teide Observatory, Instituto de Astrofisica de Canarias (IAC), in Tenerife. Gamma-ray astronomy allows the investigation of several dramatic phenomena in our Universe, such as supernova explosions or Active Galactic Nuclei and dark matter. The ASTRI MA project was conceived in 2010 by INAF to support the development of technologies whitin the Cherenkov Telescope Array (CTA) project \cite{2013APh....43....3A, 2017ICRC...35..855M}. A prototype of the ASTRI telescope \cite{pareschi2016astri} is installed in Italy at the INAF observing station located in Serra La Nave (Mt. Etna, Sicily), while the final array will consist of nine Cherenkov telescopes and cameras. 

This paper presents the LOgging UnifieD (LOUD) software architecture in the context of SCADA, the software responsible of controlling all the telescope operations. This architecture takes advantage of continuing technological evolution \cite{costa:icalepcs2019-mopha032} to respond to the challenges posed by the operation of the array, in particular to satisfy the reliability, availability and maintainability requirements of all its sub-systems and auxiliary devices. The system architecture has been designed to scale up with the number of devices to be monitored and with the number of software components to be taken into account in the distributed logging system. Internet of Things (IoT) technology allows to address the data collection from all the devices connected to the telescopes and all the other array elements.

The paper is organized as follows. Section \ref{LOUDArchitecture} introduces the architecture of the LOUD logging system; Section \ref{Data model} presents the data model and throughput; Section \ref{SystemDeployment} describes the technologies and the software deployment on which our system relies on; Section \ref{Conclusions} is for the conclusions and the future perspectives.
\section{LOUD System Architecture}\label{LOUDArchitecture}
\begin{figure}
    \centering
    \includegraphics[width=1\textwidth]{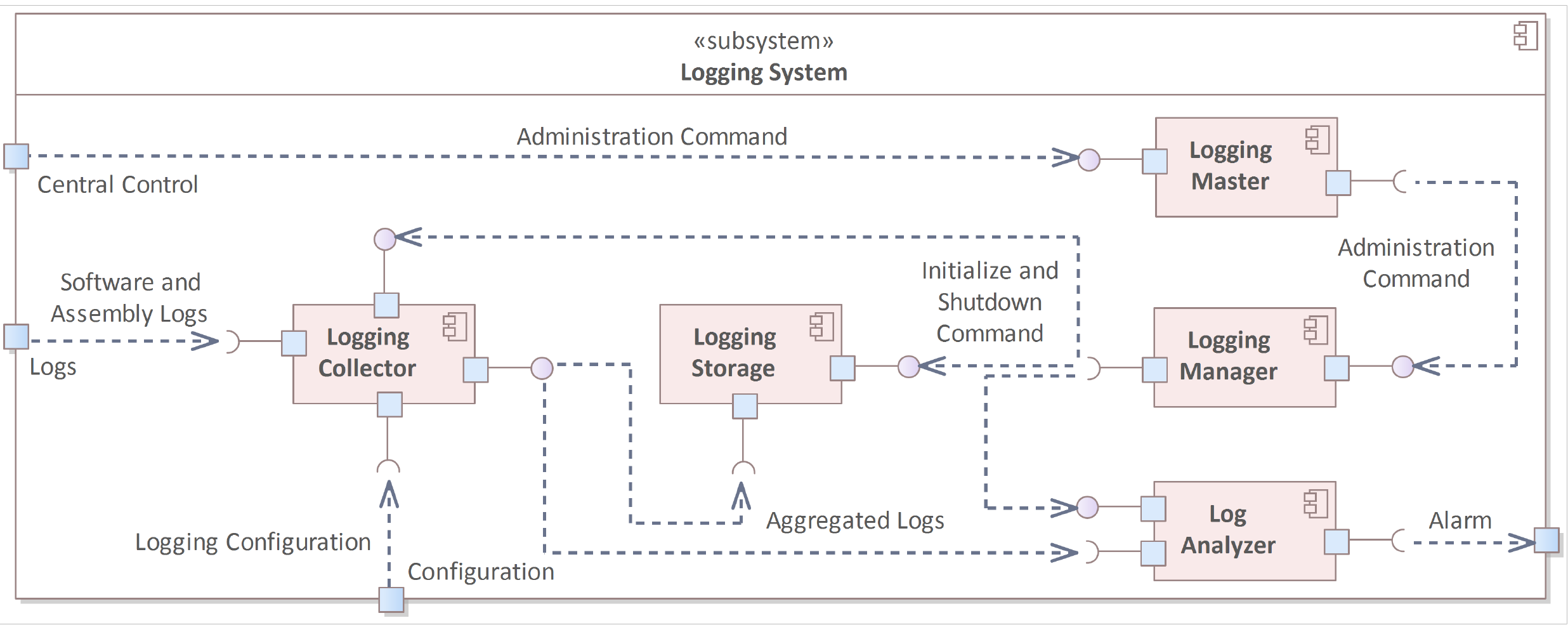}
    \caption{Logical view of LOUD. Log sources are Alma Common Software (ACS) components, OPC UA servers and low-level software.}
    \label{fig_1}
\end{figure}

\begin{figure}
    \centering
    \includegraphics[width=1\textwidth]{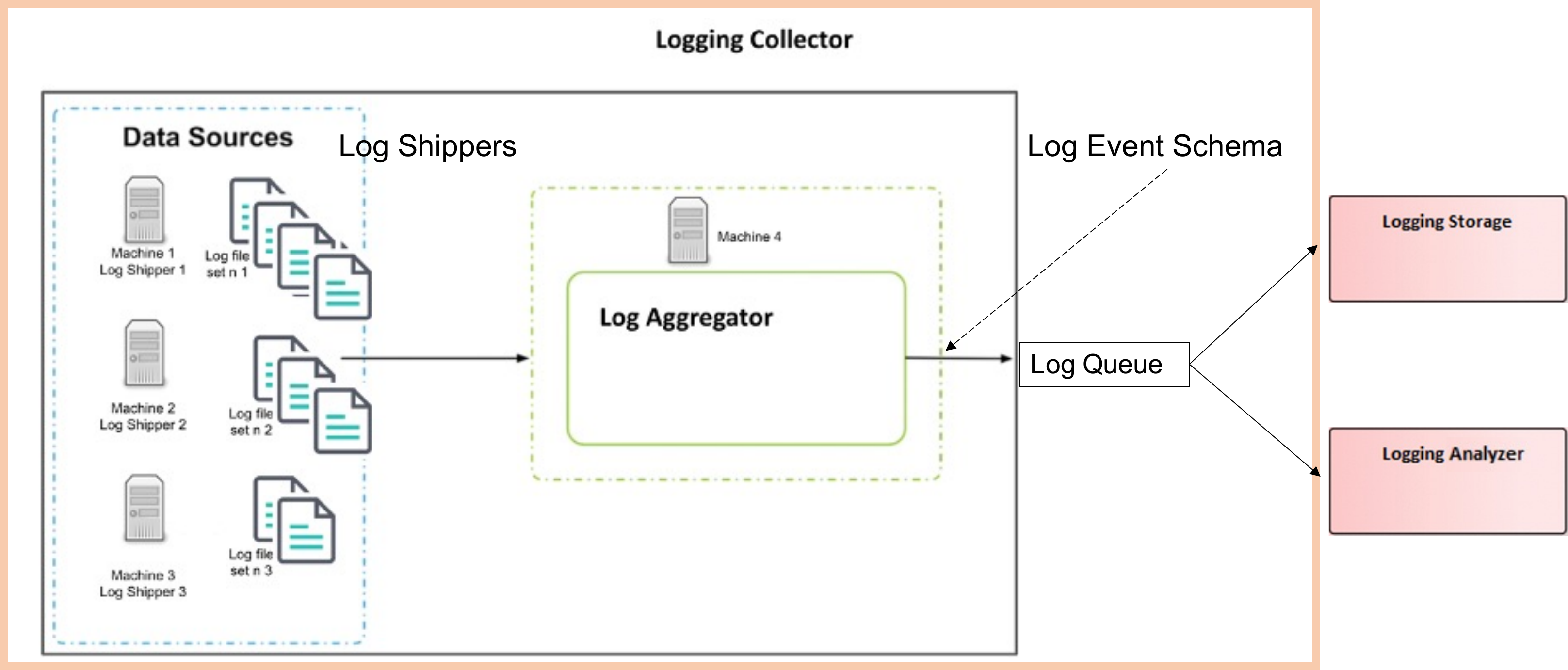}
    \caption{Architecture of the Logging Collector.}
    \label{fig_2}
\end{figure}

LOUD is responsible for acquiring logging files from all the elements of the ASTRI MA. The main building blocks of LOUD are five Logging components: Collector, Analyzer, Storage, Manager and Master. The logical architecture of the Logging System is shown in Fig. \ref{fig_1}. 
\\
The Logging Collector gets logging information from relevant software components and assemblies, which comes in three ways as described in Section \ref{Data model}. The Logging Collector is composed, in turn, of a set of Log Shippers and of an instance of Log Aggregator (Fig. \ref{fig_2}). \\
The Log Shippers are low-footprint and resource-efficient daemons that harvest a set of log files. Each Log Shipper is designed to run on a single host and gathers the set of log files produced on that machine. A Log Shipper extracts and acquires log events offering a buffer functionality and implementing back-pressure strategies towards the downstream components. Log shippers filter only those events of the required level, and send them to the Log Aggregator.\\
The Log Aggregator processes log events from Log Shippers, and sends them to Logging Storage and Logging Analyzer through a queue mechanism. A Collector based on a Log Shipper technology has been preferred over a logging library because, in a production environment, a higher reliability is provided by the mechanism of buffering implemented by a single Log Shipper. Besides, a Log Shipper implementation would be more resilient to network issues, or to congestions on the data channel implemented by the logging libraries, and it also minimizes the network load, since data is sent in batches and not individually.\\
The Logging Analyzer is responsible to analyse logging data information to trigger further alarms as well as warnings for the technical crew. \\
The Logging Storage stores logging events quantities according to desired log entry level. \\
The Logging Manager component interacts via the Logging Master with the Central Control (CC), which is the SCADA component that manages and administrates all the subsystems. It receives start-up and shut-down commands, and passes them to the LOUD systems. It provides also the LOUD status information to the CC.\\ 
The Logging Master implements and exposes to the CC a standard state machine that implements the system life cycle in a standardized way, in order to keep track of the health and the activation condition of LOUD, and to simplify the integration with other subsystems.\\
The proposed architecture is centralized, so that all log entries generated in the SCADA system can reach the central Log Aggregator. The order of the log entries is defined by their timestamp (see Section \ref{Data model}). The Logging System allows for filtering, so that log entries with insufficient priority do not get logged, whereas those with high priority get routed promptly to a dedicated high priority queue. The distributed Logging Collector allows for filtering log entries at the level of Log Shipper and Log Aggregator.
\section{Data model}\label{Data model}
The elements of the ASTRI MA are composed of hardware and application software elements. Software components, as well as SCADA, generate logs through their functioning, which record events taking place in the execution of the program in order to provide a recorded trail that can be used to understand the activity of the system and to diagnose (usually “post mortem”) problems. They are essential to understand the activities of complex systems, particularly in the case of automatized applications with little user interaction. The software logs are archived for the usage of the software maintenance.\\
LOUD provides a standardized infrastructure for logging data sources via software within the array element control systems. This ensures the required modularity, flexibility, and performance since logging is, in general, a 24/7 activity that shall occur any time the corresponding software process is running.

In this section, we describe the logging data exchange of the interface between the LOUD system (the target system) and a software application contained in a generic element of ASTRI MA (the source system). \\
The logging information is transferred from the data source components to SCADA in three ways:
\begin{itemize}
    \item[] Alma Common Software (ACS): ACS components use the standard ACS logging mechanism to insert logs \cite{chiozzi2004alma}. LOUD is able to catch these logs and to drive them into its pipeline; 
    \item[] OPC UA: OPC UA servers use the OPC UA logging infrastructure \cite{OPCUA}. Logs are stored in LOUD by placing log entries in files that are stored in a folder structure;
    \item[] Low-level software logs: Low-level software processes, which are not using ACS or OPC UA, store logs in the same way as OPC UA logs.
\end{itemize}
Irrespective of the logging mechanism used, the elemets of ASTRI MA insert low-level logs by piping log data into files located in the on-site data centre.\\
The volume of the logs produced by the software components is tuned properly in order to avoid generating excessive logging information during regular operations, but, also, to provide enough information during debugging campaigns. This is achieved by means of configurable filtering based on logging levels, as described in Section \ref{LOUDArchitecture}.\\
Log files used for OPC UA and low-level software are plain ASCII files UTF-8 encoding with extension "\textit{.log}". The file name follows the naming scheme:\\\\
\textit{componentInstanceName\_<YYYY>-<MM>-<DD>.log},\\\\
for the currently active log file, and:\\\\
\textit{componentInstanceName\_<YYYY>-<MM>-<DD>.<num>.log},\\\\
for files generated earlier in the same folder, where: \textit{componentInstanceName} is the name of the instance of the component creating the logs, \textit{<YYYY>-<MM>-<DD>} is the UTC date corresponding to the instant when the file is opened, and \textit{<num>} is an integer number starting from 1 for the first log file created during the day, and increasing monotonically for additional log files created the same day.\\
A software logging file contains one line per new log entry, which will follow this structure: \\\\
\textit{<sourceTimestamp> <loggingLevel> <file> <line> <routine> <sourceObject> <logAudience> <Message>}.\\\\
The \textit{<sourceTimestamp>} field specify the UTC time when the log file was generated, following the ISOT (ISO 8601) time format: \\\\
\textit{YY-MM-DDTHH:mm:ss},\\\\
where \textit{YY} is year, \textit{MM} is month, \textit{DD} is day, \textit{HH} is hour, \textit{mm} is minute, and \textit{ss} is second with a precision to one millisecond. \\
The \textit{<loggingLevel>} field describes the level of the log entry in term of priority according to the categorization defined by SCADA, in the following way:
\begin{itemize}
    \item[] TRACE: generated whenever a function is entered, and used to report calls to a function;
    \item[] DEBUG: used only for system debugging;
    \item[] DELOUSE: provides the highest level of detail for debugging the system;
    \item[] INFO: used to publish information of interest during the normal operation of the system;
    \item[] NOTICE: useful for logging normal but significant activity of the system, for example startup or shutdown of individual services. They denote important situations in the system, but not necessarily error/fault conditions;
    \item[] WARN: used to report to conditions that are not errors but that could lead to errors/problems;
    \item[] ERROR: denotes error conditions;
    \item[] CRITICAL: indicates an Alarm condition that shall be reported to the operators through the Human Machine Interface (HMI);
    \item[] ALERT: denotes an Alarm condition that shall be reported to the operators through the HMI. This indicates a problem more important than Critical;
    \item[] EMERGENCY: denotes an Alarm condition of the highest priority.
\end{itemize}
The \textit{<file>}, \textit{<line>} and \textit{<routine>} fields are optional and describe, respectively, the identification of the source file, the line number in the source code where the log entry was submitted, and the name of the function where the log entry was submitted from. \\
\textit{<sourceObject>} and \textit{<logAudience>} are the name of the process from which the log entry was generated and the audience of this log, respectively, while \textit{<message>} is the real log message, which must include meaningful information that is relevant for the log audience. 
\subsection{Logging data throughput}\label{Data throughput}
At any given time, all the logs produced by all the applications belonging to an element of the ASTRI MA will not produce a data volume higher than 10 MB/s, including, in the case of a telescope, all the log files produced by the applications associated to the telescope itself and to the host structure. This value does not consider any kind of compression or later filtering in SCADA. \\
In  the  framework  of ASTRI SCADA, for storing log information, we expect a maximum data rate in input of about 200 Mbps.
\section{System Deployment}\label{SystemDeployment}
The LOUD system is written in Java programming language and it is integrated with the ALMA Common Software (ACS), an open-source framework on which relies the software operating the ALMA observatory. Since ACS allows the usage of Java, C++ and Python programming languages, our system is able to accept logging API written in any one of these languages. \\
We build and manage our Java-based system in an automated way, through Apache Maven \cite{miller2010apache}.\\
To exchange the acquired data among the heterogeneous SCADA subsystems, we use Apache Avro \cite{Avro}, a data serialization framework that, in turn, uses JSON \cite{pezoa2016foundations} for defining schemas for information transmission.\\ 
We use Apache Kafka \cite{kafka} to manage the data flow. Kafka is a distributed event streaming platform designed to handle data streams from multiple sources and deliver them to multiple consumers.\\
To forward and centralize logs generated by SCADA, we use a set of distributed lightweight Log Shippers (see Section \ref{LOUDArchitecture}) based on Elastic Filebeat \cite{Filebeat}. Those log events are ingested, filtered and manipulated by a centralized Log Aggregator (Section \ref{LOUDArchitecture}) based on Elastic Logstash \cite{Logstash}, which acts as a data processing pipeline that, in turn, sends them to Kafka.\\
We exploit Apache Cassandra \cite{CassandraPaper, Cassandra} as our database management system (DBMS), which is specifically designed to handle large amounts of data \cite{Abramova2013}.\\
We make use of the Docker platform \cite{merkel2014docker} that provides the ability to package and run an application in a loosely isolated environment called container.\\
Finally, to easily distribute, replicate and scale our containerized applications we exploit the Kubernetes orchestrator \cite{Kubernetes}.\\
\subsection{Log Simulator}
To test our architecture and software deployment, we have developed a Log Simulator in Python. \\
The simulator is able to produce a number of log files in parallel, in the expected format. The log entries are randomly chosen from a lookup table, a small database obtained by real log data. The values are sampled with a configurable probability distribution based on the log level. In this way, it is possible to obtain a series of log files whose log level entries are distributed according to the purpose of the simulation. This is useful in the test phase since it allows the user to reproduce several behaviours of the logging production, e.g. in normal working or in ad hoc conditions. Besides, the simulator is able to generate log entries with a desired production rate for each log file.
\section{Conclusion and Future Perspectives}\label{Conclusions}
We presented the architecture of the LOgging UnifieD (LOUD) system that is responsible for gathering, filtering, exposing and storing logs data, which are needed to record the operational activities of a telescope array. LOUD was designed and built exploiting the current most advanced technologies in the field of the Internet of Things, and it is based on open source software. In the future, we plan to integrate Deep Learning algorithms to perform anomaly detection and failure prediction based on the log events.
\acknowledgments
This work was conducted in the context of the ASTRI Project. This work is supported by the Italian Ministry
of Education, University, and Research (MIUR) with funds specifically assigned to the Italian National Institute of Astrophysics (INAF). We acknowledge support from the Brazilian Funding Agency FAPESP (Grant 2013/10559-5) and from the South African Department of Science and Technology through Funding Agreement 0227/2014 for the South African Gamma-Ray Astronomy Programme. IAC is supported by the Spanish Ministry of Science and Innovation (MICIU).
%
%
% References
\bibliographystyle{JHEP}
\bibliography{main} % bibliography data in report.bib

\providecommand{\href}[2]{#2}\begingroup\raggedright\begin{thebibliography}{10}

\bibitem{antonelliLA}
L.A.~Antonelli, \emph{The astri mini-array at teide observatory},  in
  \emph{37th International Cosmic Ray Conference (ICRC2021)}, vol.~This
  Proceeding of \emph{International Cosmic Ray Conference Series}, 2021.

\bibitem{2013APh....43....3A}
{{Cherenkov Telescope Array Consortium}}, \emph{{Introducing the CTA concept}},
  \href{https://doi.org/10.1016/j.astropartphys.2013.01.007}{\emph{Astroparticle
  Physics} {\bfseries 43} (2013) 3}.

\bibitem{2017ICRC...35..855M}
M.C.~{Maccarone}, \emph{{ASTRI for the Cherenkov Telescope Array}},  in
  \emph{35th International Cosmic Ray Conference (ICRC2017)}, vol.~301 of
  \emph{International Cosmic Ray Conference}, p.~855, Jan., 2017
  [\href{https://arxiv.org/abs/1709.03078}{{\ttfamily 1709.03078}}].

\bibitem{pareschi2016astri}
G.~Pareschi, \emph{The astri sst-2m prototype and mini-array for the cherenkov
  telescope array (cta)},  in \emph{Ground-based and Airborne Telescopes VI},
  vol.~9906, p.~99065T, International Society for Optics and Photonics, 2016.

\bibitem{costa:icalepcs2019-mopha032}
A.~Costa et~al., \emph{{Big Data Architectures for Logging and Monitoring Large
  Scale Telescope Arrays}},  in \emph{Proc. ICALEPCS'19}, no.~17 in
  International Conference on Accelerator and Large Experimental Physics
  Control Systems, pp.~268--271, JACoW Publishing, Geneva, Switzerland, 08,
  2020, \href{https://doi.org/10.18429/JACoW-ICALEPCS2019-MOPHA032}{DOI}.

\bibitem{chiozzi2004alma}
G.~Chiozzi et~al., \emph{The alma common software: a developer-friendly
  corba-based framework},  in \emph{Advanced Software, Control, and
  Communication Systems for Astronomy}, vol.~5496, pp.~205--218, International
  Society for Optics and Photonics, 2004.

\bibitem{OPCUA}
``Opc ua online reference.''
  \url{https://reference.opcfoundation.org/v104/Core/docs/Part4/}.

\bibitem{miller2010apache}
F.P.~Miller, A.F.~Vandome and J.~McBrewster, \emph{Apache Maven}, Alpha Press
  (2010).

\bibitem{Avro}
``Apache avro.'' \url{http://avro.apache.org}.

\bibitem{pezoa2016foundations}
F.~Pezoa, J.L.~Reutter, F.~Suarez, M.~Ugarte and D.~Vrgo{\v{c}},
  \emph{Foundations of json schema},  in \emph{Proceedings of the 25th
  International Conference on World Wide Web}, pp.~263--273, International
  World Wide Web Conferences Steering Committee, 2016.

\bibitem{kafka}
N.~Garg, \emph{Apache Kafka}, Packt Publishing (2013).

\bibitem{Filebeat}
``Elastic filebeat.'' \url{https://www.elastic.co/beats/filebeat}.

\bibitem{Logstash}
``Elastic logstash.'' \url{https://www.elastic.co/logstash}.

\bibitem{CassandraPaper}
A.~Lakshman and P.~Malik, \emph{Cassandra: A decentralized structured storage
  system}, \href{https://doi.org/10.1145/1773912.1773922}{\emph{SIGOPS Oper.
  Syst. Rev.} {\bfseries 44} (2010) 35–40}.

\bibitem{Cassandra}
``Apache cassandra.'' \url{https://cassandra.apache.org}.

\bibitem{Abramova2013}
V.~Abramova and J.~Bernardino, \emph{Nosql databases: Mongodb vs cassandra},
  {\emph{C3S2E '13: Proceedings of the International C* Conference on Computer
  Science and Software Engineering} (2013) }.

\bibitem{merkel2014docker}
D.~Merkel, \emph{Docker: lightweight linux containers for consistent
  development and deployment}, {\emph{Linux journal} {\bfseries 2014} (2014)
  2}.

\bibitem{Kubernetes}
``Kubernetes.'' \url{https://kubernetes.io}.

\end{thebibliography}\endgroup
\end{document}